\documentclass{aastex63}
\usepackage{amsmath,amssymb,bbold,mathrsfs}
\usepackage{textcase}
\usepackage{subfigure}

\usepackage{xcolor}
\usepackage{soul}
\definecolor{g}{RGB}{0,160,0}
\setstcolor{g}
\definecolor{b}{RGB}{0,0,160}
\setstcolor{b}
\definecolor{r}{RGB}{160,0,160}
\setstcolor{r}

\newcommand{\TAUE}[1]{${\rm log_{10}}(\tau_{500})=#1$}
\newcommand{\TAUA}[1]{${\rm log_{10}}(\tau_{500})\approx#1$}

\newcommand{\blosp}{$B_{\rm \parallel}$}
\newcommand{\blos}{$B_{\rm \parallel}$}
\newcommand{\vlosp}{$v_{\rm \parallel}$}
\newcommand{\vlos}{$v_{\rm \parallel}$}
\newcommand{\vmip}{$v_{\rm turb}$}
\newcommand{\vmi}{$v_{\rm turb}$}
\newcommand{\Pgp}{$P_{\rm g}$}
\newcommand{\Pg}{$P_{\rm g}$}
\newcommand{\fref}[1]{Fig.~\ref{#1}}
\newcommand{\sref}[1]{Section~\ref{#1}}

\newcommand{\bzp}{$B_z$}
\newcommand{\bz}{$B_z$}
\newcommand{\vzp}{$v_z$}
\newcommand{\vz}{$v_z$}

\usepackage[nonumberlist,nosuper]{glossaries}
\setacronymstyle{long-short}
\newacronym{los}{LOS}{line of sight}
\newacronym{clasp2}{CLASP2}{Chromospheric LAyer Spectropolarimeter}
\newacronym{mo}{MO}{magneto-optical}
\newacronym{wfa}{WFA}{weak field approximation}
\newacronym{psf}{PSF}{point spread function}
\newacronym{fwhm}{FWHM}{full width half maximum}
\newacronym{t}{$T$}{temperature}
\newacronym{pg}{\Pgp}{gas pressure}
\newacronym{vmi}{\vmip}{microturbulent velocity}
\newacronym{blos}{\blosp}{magnetic field longitudinal component}
\newacronym{vlos}{\vlosp}{velocity longitudinal component}
\newacronym{bz}{\bzp}{vertical component of the magnetic field}
\newacronym{vz}{\vzp}{vertical component of the macroscopic velocity}
\newacronym{mhd}{MHD}{magneto-hydrodynamic}
\newacronym{prd}{PRD}{partial frequency redistribution}
\begin{document}

\title{Tomography of a solar plage with the Tenerife Inversion Code}
\shorttitle{Solar plage tomography with TIC}

\author[0000-0001-5612-4457]{Hao Li}
\affil{Instituto de Astrof\'{\i}sica de Canarias, E-38205 La Laguna, Tenerife, Spain}
\affil{Departamento de Astrof\'\i sica, Universidad de La Laguna, E-38206 La Laguna, Tenerife, Spain}
\author[0000-0003-1465-5692]{T.\ del Pino Alem\'an}
\affil{Instituto de Astrof\'{\i}sica de Canarias, E-38205 La Laguna, Tenerife, Spain}
\affil{Departamento de Astrof\'\i sica, Universidad de La Laguna, E-38206 La Laguna, Tenerife, Spain}
\author[0000-0001-5131-4139]{J.\ Trujillo Bueno}
\affil{Instituto de Astrof\'{\i}sica de Canarias, E-38205 La Laguna, Tenerife, Spain}
\affil{Departamento de Astrof\'\i sica, Universidad de La Laguna, E-38206 La Laguna, Tenerife, Spain}
\affil{Consejo Superior de Investigaciones Cient\'{\i}ficas, Spain}
\author[0000-0001-8830-0769]{R.\ Ishikawa}
\affil{National Astronomical Observatory of Japan, Mitaka, Tokyo 181-8588, Japan}
\author[0000-0001-9095-9685]{E.\ Alsina Ballester}
\affil{Instituto de Astrof\'{\i}sica de Canarias, E-38205 La Laguna, Tenerife, Spain}
\affil{Departamento de Astrof\'\i sica, Universidad de La Laguna, E-38206 La Laguna, Tenerife, Spain}
\author[0000-0002-9921-7757]{David E. McKenzie}
\affiliation{NASA Marshall Space Flight Center, Huntsville, AL 35812, USA}
\author[0000-0003-0972-7022]{Fr\'ed\'eric Auch\`ere}
\affiliation{Institut d'Astrophysique Spatiale, 91405 Orsay Cedex, France}
\author[0000-0003-1057-7113]{Ken Kobayashi}
\affiliation{NASA Marshall Space Flight Center, Huntsville, AL 35812, USA}
\author[0000-0003-3765-1774]{Takenori J. Okamoto}
\affil{National Astronomical Observatory of Japan, Mitaka, Tokyo 181-8588, Japan}
\author[0000-0002-3770-009X]{Laurel A. Rachmeler}
\affiliation{National Oceanic and Atmospheric Administration, \\ 
National Centers for Environmental Information, Boulder, CO 80305, USA}
\author[0000-0003-3034-8406]{Donguk Song}
\affiliation{Korea Astronomy and Space Science Institute 776, \\
Daedeokdae-ro, Yuseong-gu, Daejeon 305-348, Republic of Korea}
\affil{National Astronomical Observatory of Japan, Mitaka, Tokyo 181-8588, Japan}

%###############################################################################
%###############################################################################
%###############################################################################

\begin{abstract}
We apply the Tenerife Inversion Code (TIC) to the plage spectropolarimetric
observations obtained by the Chromospheric LAyer SpectroPolarimeter (CLASP2).
These unprecedented data consist of full Stokes 
profiles in the spectral region around the \ion{Mg}{2} h and k lines for a single slit
position, with around two thirds of the 200~arcsec slit crossing a plage region
and the rest crossing an enhanced network. A former analysis of these data 
had allowed us to infer the longitudinal component of the magnetic field by
applying the weak field approximation (WFA) to the circular polarization profiles, 
and to assign the inferred magnetic fields to different layers of the solar atmosphere
based on the results of previous theoretical radiative transfer investigations. 
In this work, we apply the recently developed TIC to the same data. We obtain the
stratified model atmosphere that fits the intensity and circular polarization
profiles at each position along the spectrograph slit and we compare our results
for the longitudinal component of the magnetic field with the previously obtained
WFA results, highlighting the generally good agreement in spite of the fact that
the WFA is known to produce an underestimation when applied to the outer lobes
of the \ion{Mg}{2} h and k circular polarization profiles. Finally, we use the
inverted model atmospheres to give a rough estimation of the energy that could
be carried by Alfv\'en waves propagating along the chromosphere in the plage
and network regions, showing that it is sufficient to compensate the estimated
energy losses in the chromosphere of solar active regions.
\end{abstract}

%###############################################################################
%###############################################################################
%###############################################################################

\section{Introduction} \label{S-intro}

One of the major challenges in solar physics is the determination of the
magnetic field vector in the solar chromosphere 
\citep[e.g., the review by][]{JTB-TdPA-ARAA}, which in turn is key to
understanding the physics of this interface region between the
underlying photosphere and the overlying corona. In the chromosphere the magnetic
pressure starts to overcome the gas pressure and the magnetic field increasingly
dominates the plasma dynamics and structure
\citep[e.g., the review by][and references therein]{Carlsson2019ARA&A}.

Most of the spectral lines that form in the upper solar chromosphere and in the
transition region up to the million-degree corona are resonance lines in the
ultraviolet (UV), a spectral region not accessible by ground based facilities. 
The polarization signals of UV spectral lines encode information about the plasma
magnetic field in such key regions of the solar upper atmosphere
\citep{Trujillo2017SSRv}.

Motivated by theoretical investigations of the polarization in the \ion{Mg}{2}
h-k doublet around 280~nm \citep{Belluzzi2012ApJ}, the CLASP2
\citep[Chromospheric LAyer SpectroPolarimeter;][]{Narukage2016} suborbital space
experiment was performed in 2019, allowing the observation of the intensity and
polarization in this near-UV region of the spectrum in both a quiet Sun region
near the limb \citep{Rachmeler2022} and in an active region plage
\citep{Ishikawa2021}. The CLASP2 mission was a true success and it achieved
unprecedented observations of the UV solar spectrum between approximately
279.30 and 280.68~nm, which includes the \ion{Mg}{2} h and k resonance lines and
the three resonance lines of \ion{Mn}{1} at 279.56, 279.91, and 280.19~nm.

The observations corresponding to the plage region showed fractional
circular polarization signals of significant amplitude in the
above-mentioned \ion{Mg}{2} and \ion{Mn}{1} spectral lines. A first analysis
of these data enabled an estimate of the longitudinal component of the magnetic
field by applying the \gls*{wfa} to the intensity and circular polarization profiles
\citep{Ishikawa2021}. Based on theoretical
radiative transfer investigations, the magnetic field strengths inferred from
the \ion{Mg}{2} h-k doublet \citep{Tanausu2020ApJ,Afonso2023ApJ} and \ion{Mn}{1} triplet
\citep{Tanausu2022ApJ} were assigned to different layers of the solar
atmosphere. In particular, the \ion{Mg}{2} lines provided the longitudinal
component of the magnetic field in the upper and middle chromosphere,
while the \ion{Mn}{1} lines provided it in the lower chromosphere.
Together with the simultaneous observations carried out by HINODE/SOT-SP in
photospheric lines, this resulted in a map of the magnetic field (its longitudinal
component) from the photosphere to the base of the corona \citep{Ishikawa2021}.

Recently, \cite{Li2022} developed the Tenerife Inversion Code (TIC), a non-LTE
inversion code of Stokes profiles produced by the scattering of anisotropic
radiation and the Hanle and Zeeman effects in spectral lines.
The TIC takes into account the effects of radiative 
transfer in one-dimensional models of the solar atmosphere, as well as 
the effects of partial frequency redistribution and $J$-state interference 
(necessary to model the \ion{Mg}{2} h-k doublet) and of the radiation field
anisotropy (which has a significant impact even on 
the circular polarization of the \ion{Mg}{2} h-k doublet; \citealt{AlsinaBallester2016ApJ,Tanausu2016ApJ,Li2022}). 
The TIC can be applied to a variety of spectropolarimetric observations in order to get
the stratification of the magnetic field in the observed region, including the   
spectropolarimetric data of the \ion{Mg}{2} h and k lines obtained by CLASP2.

In this work, we apply TIC to the Stokes $I$ and $V$ profiles of the \ion{Mg}{2} h
and k lines and of the two blended lines of the \ion{Mg}{2} subordinate triplet
at $279.88$~nm observed by CLASP2 in an active region plage. 
The Stokes Q and U signals were also successfully measured by CLASP2 in the 
observed active region plage (see figure S2 of \citealp{Ishikawa2021}) and in the 
observed quiet region close to the limb (see figure 2 of \citealp{Rachmeler2022} and 
figure 6 of \citealp{JTB-TdPA-ARAA}), however, the inversion of the four Stokes 
parameters is significantly more complex, both in terms of computational time and from 
the physics involved in the scattering problem, and therefore is left for a future investigation.
In \sref{S-info} we
briefly describe the most relevant aspects of the data as well as our inversion
strategy. In \sref{S-results} we show the results of the inversion. We compare the
resulting  longitudinal component of the magnetic field with the \gls*{wfa} inference
by \cite{Ishikawa2021}. We then use the inverted model atmosphere in order to estimate
the energy flux that could be carried by Alfv\'en waves propagating through the
chromospheres of plage and network regions. Finally, we summarize our conclusions in
\sref{S-conclusions}. 

%###############################################################################
%###############################################################################
%###############################################################################

\section{Observations and inversion strategy}\label{S-info}

In this section we summarize the main characteristics of the observations
and describe the inversion strategy we have applied with the TIC.

%###############################################################################

\subsection{Observation}\label{SS-data}

The subject data of this work was obtained by the CLASP2
launched by a NASA sounding rocket on 11 April 2019. In particular, here we
focus on the data corresponding to the plage target. The first analysis of this
data was carried out by \cite{Ishikawa2021}, who applied the  
\gls*{wfa} to the intensity and circular
polarization profiles of the \ion{Mg}{2} h-k doublet and \ion{Mn}{1} lines
at $279.91$ and $280.19$~nm to infer the longitudinal component of the
magnetic field in different layers of the solar atmosphere.

The data used in this work are the Stokes $I$ and $V$ profiles 
at each point of the 196~arcsec slit of the spectrograph covering
an active region plage and its surrounding enhanced network at the East side
of NOAA~12738 (see panel C of \fref{fig1}).
The data was recorded between 16:53:40 and 16:56:16~UT.  
($\approx2.5$~min). The spectral range covers from 279.30 to 280.68~nm with
a spectral sampling of 49.9~m\AA/pixel. The instrument's spectral profile
can be approximated with a Gaussian with a full width at half maximum (FWHM)
of 110~m$\rm \AA$, which results from the convolution of the spectrograph's slit
with the spectral point spread function \citep[PSF;][]{Song2018SPIE,Tsuzuki2020SPIE}. 
The spectral resolution of the data is, therefore, about 28000. 
The \ion{Mg}{2} h and k lines at 279.6 and 280.3~nm, respectively, as well as the
two subordinate lines at 279.88~nm (they are blended), are found in this spectral
range. 

The Stokes profiles of the \ion{Mg}{2} h-k doublet show sizable circular
polarization in the bright plage and weaker circulation polarization in
the enhanced network region, from which we can infer the longitudinal component
of the magnetic field in the solar chromosphere. 

%###############################################################################

\subsection{Inversion strategy}\label{SS-inv}

We apply TIC in order to infer the stratification of the thermodynamical
parameters\textemdash namely \gls*{t}, \gls*{pg}, \gls*{vmi}, the \gls*{vz}, and
the \gls*{bz}\textemdash in the solar plage target observed by the CLASP2. 
TIC is based on the HanleRT forward engine
\citep{Tanausu2016ApJ,Tanausu2020ApJ}, which implements the coherent
scattering formalism of \cite{Casini2014ApJ,Casini2017ApJ,Casini2017bApJ}. 
At each pixel along the spatial direction of the spectrograph's slit  
we invert the observed Stokes profiles, after temporally co-adding the data to attain a suitable
signal to noise ratio for the inversion.

We use a 6-levels atomic model (5 \ion{Mg}{2} levels and the \ion{Mg}{3} ground level), 
including the h-k doublet and the subordinate lines in the observed range
\citep[see also][]{Tanausu2020ApJ}.
The stratification in the model atmosphere is sampled with 60 nodes between
\TAUE{-8} and 1. The nodes are not equally spaced in order to refine the sampling
between \TAUE{-6.5} and -3, which corresponds to the region of formation of 
the \ion{Mg}{2} h and k lines. We invert the Stokes profiles in 5 cycles:

\begin{itemize}
  \item[1.] Invert $T$, {\vmi}, and {\vz} with 4, 2, and 3 nodes, respectively,
            using only the intensity profile. The initial model
            atmosphere is the static model C of \cite{Fontenla1993}.
  \item[2.] Invert $T$, {\vmi}, and {\vz} with 7, 4, and 5 nodes, respectively,
            using only the intensity profile.
  \item[3.] Invert {\bz} with 1 node,
  	    using both Stokes I and V, and neglecting the anisotropy of the radiation field.
  \item[4.] Invert $T$, {\vmi}, {\vz}, and {\bz} with 7, 4, 5, and 5 nodes,
            respectively, using both Stokes I and V, and neglecting the anisotropy of the radiation field.
  \item[5.] Invert {\bz} with 5 nodes, using both Stokes I and V, 
  	   and taking into account the radiation field anisotropy.
\end{itemize}

Every cycle after the first is initialized from the model atmosphere resulting
from the previous cycle. The first two cycles results in the model atmosphere
that fits the intensity profile observed at the spatial pixel under consideration, and 
they are relatively fast 
because polarization is not taken
into account. Note that because we assume hydrostatic equilibrium, {\Pg} is fully
determined by its value at the top boundary and the temperature stratification.
Whenever the thermodynamic variables are being inverted, the boundary value of
{\Pg} also is. The second cycle is more time consuming due to the increased
number of nodes, but can provide added complexity (if needed) to the
stratification of thermodynamic parameters. The radiation field anisotropy has
a significant impact on the circular polarization of the \ion{Mg}{2} h-k doublet
\citep{AlsinaBallester2016ApJ,Tanausu2016ApJ,Li2022} and it is necessary to take it
into account.
However, accounting for the anisotropy makes the inversion appreciably
slower. For this reason the first two magnetic cycles (3 and 4) neglect this
contribution. The relatively fast third cycle gives a very rough estimation of {\bzp}, and
then the fourth cycle adds stratification to {\bz} while also letting the thermodynamic
parameters slightly change to improve the simultaneous fit of both intensity and
circular polarization. Finally, in the fifth cycle the radiation field anisotropy
is included and {\bz} is readjusted.

It is important to clarify that the \gls*{los} is not at the disk center 
(it varies from $\mu\approx0.65$ to $\mu\approx0.85$ along the slit). 
When including the radiation field
anisotropy, the geometry with respect to the propagation directions of the radiation
within the atmosphere is important, and not just along the \gls*{los}. If 
the macroscopic velocity or the magnetic field vector are not vertical, the forward problem is
no longer axially symmetric and the computing time is significantly increased.
This is the reason why the inversion is on {\vz} and {\bzp}. These vertical components
are then projected onto the \gls*{los}, and thus we plot the result for the
\gls*{vlos} and the \gls*{blos}. In the inversions shown in this work we assume
that a single magnetic field fills the whole pixel (i.e., we assume the magnetic filling
factor to be unity).

%###############################################################################
%###############################################################################
%###############################################################################

\section{Results}\label{S-results}

In this section we show the results of the inversion of the intensity and
circular polarization profiles obtained by the CLASP2 in the observed active
region plage
(see \sref{SS-data}). We compare our results with those of
\cite{Ishikawa2021}, which inferred {\blos} at different heights in the solar
atmosphere by applying the \gls*{wfa}. Finally, we use the inferred atmospheric
parameters to estimate the sound speed, the Alfv\'en velocity, the plasma $\beta$  
parameter, as well as the energy flux of Alfv\'en waves.

%###############################################################################

\subsection{Inverted model atmosphere}\label{SS-Rinv}

\begin{figure*}[htp]
\center
\includegraphics[width=0.90\textwidth]{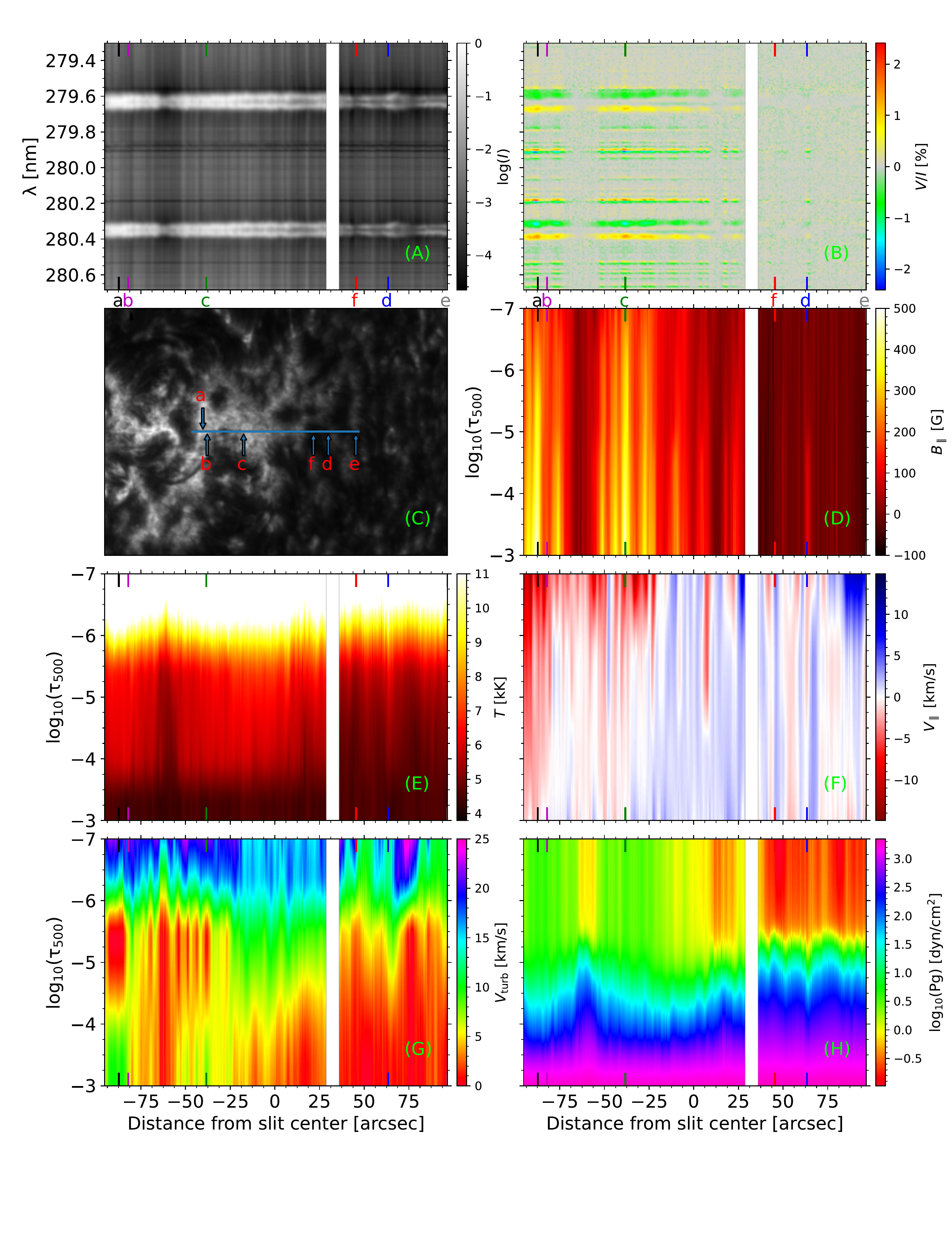}
\vspace{-1cm}
\caption{Intensity profiles (panel A) normalized to 
	the maximum, in log scale, and fractional
         circular polarization $V/I$ profiles (panel B), measured by CLASP2. 
         Panel (C): slit-jaw image of Lyman$-\alpha$ recorded by CLASP2. The blue line denotes the
         location of the spectrograph's slit.
         Panels D to H show the results for {\blos}, $T$, {\vlos}, {\vmi}, and {\Pg}, respectively. 
         The a, b, c, d, e, and f labels indicate the
         locations of the profiles shown in \fref{fig2}.
          }
\label{fig1}
\end{figure*}

Following the strategy in \sref{SS-inv}, we apply the TIC code to 
the Stokes profiles for each 
slit position of the CLASP2 plage target observation (\sref{SS-data}). The inversion
results are shown in \fref{fig1}.

The intensity shows the typical two-peaks profiles and they are brighter
inside the plage (located approximately between slit positions at -95 and 28~arcsec)
than in the surrounding enhanced network regions (see panel A of \fref{fig1}).
The white vertical band at around 34~arcsec corresponds to the lack of data due
to dust in the instrument. It is well known that the h and k line centers (often
dubbed h$_3$ and k$_3$, respectively) form in the upper chromosphere, while the peaks
(dubbed h$_{2v}$ and h$_{2r}$, for the h line, and k$_{2v}$ and k$_{2r}$ for the k line)
form in the middle chromosphere, and the minima outside the peaks (dubbed h$_1$ and k$_1$,
respectively) form close to the temperature minimum of standard semi-empirical models 
\citep{Vernazza1981ApJS}. The \ion{Mg}{2} h and k lines are thus sensitive to a relatively
wide range of heights through the solar chromosphere.

The $V/I$ profiles are shown in panel B. These signals are the strongest inside
the plage region, as expected, resulting in larger {\blos}. In this region, 
the strengths of the magnetic flux concentrations 
reach around $300$~G in the upper
chromosphere (from \TAUA{-5.5} to $\approx-6.5$, see panel D of \fref{fig1}). 
The plage region also shows higher temperatures at chromospheric heights 
(from \TAUA{-3.5} to $\approx-6.0$) with respect to the surrounding enhanced network
region (see panel E of \fref{fig1}).

The {\vlos} macroscopic velocity gradients are necessary to get the observed asymmetries between k$_{2v}$
and k$_{2r}$, and equivalently for the h line \citep{Leenaarts2013ApJ}, as
can be seen in panel F of \fref{fig1}.\footnote{The CLASP2 data was calibrated in
wavelength by assuming that the \ion{Mn}{1} lines were at rest. The solar rotation was then
subtracted \citep{Snodgrass1990ApJ}.} The inverted macroscopic velocity is mostly
downward (red color in the figure) at the slit positions between approximately -95 and
-25~arcsec, while mostly upward (blue color in the figure) between approximately -25 and
30~arcsec.

The {\vmi} parameter accounts mainly for the missing broadening due to the
three-dimensional plasma dynamics. We get relatively small values up to  
the mid chromosphere (\TAUA{-5.8}) in most of the slit, significantly increasing in
the upper chromosphere (see panel G of \fref{fig1}). Generally, {\vmi} is larger 
in the bright plage region.

The stratification of {\Pg} is calculated assuming hydrostatic equilibrium
\citep[e.g.,][]{Mihalas1970}, although its value in the upper boundary is
allowed to change in the inversion (see panel H of \fref{fig1}). The
electron and hydrogen number densities are computed in the forward solver by
assuming LTE and solving the equation of state with the method of
\citet{Wittmann1974solphys}.
 
\begin{figure*}[htp]
\center
\includegraphics[width=0.9\textwidth]{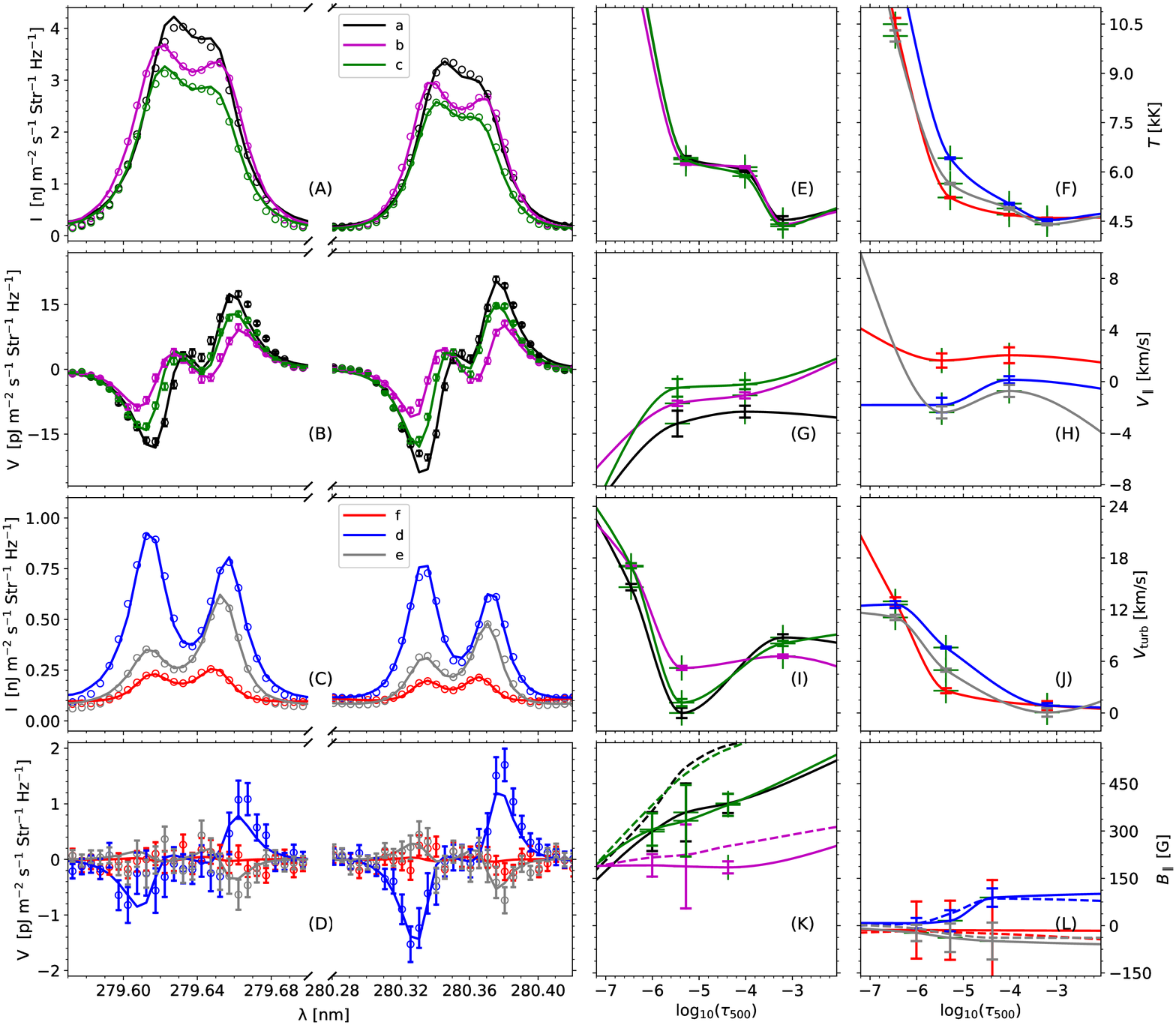}
\vspace{-1.5cm}
\caption{Observed (open circles) and fit (solid curves) profiles (A, B,
C, and D) for the intensity (A and C) and circular polarization
(B and D) at the slit locations indicated in \fref{fig1} in
the plage (panels A and B; locations a, b, and c) and in the enhanced
network and quiet sun (panels C and D; locations d and e and f,
respectively). The two left columns show the
k and h lines. The error bars in the circular polarization profiles
indicate the photon noise.
Fit profiles with the full spectral range, including the
subordinate lines, are shown in \fref{figa1} for the plage (b) and
enhanced network (d) slit locations.
The two right columns show, from top to bottom, the inverted
T (E, F), {\vlos} (G, H), {\vmi} (I, J), and {\blos} (K, L) stratification, with
the color of each curve corresponding to a slit position as
indicated in the legends in panels A and C. In panels K and L,
the solid (dashed) curves show the result from the fifth (fourth) inversion
cycle, that is, including (neglecting) the radiation field
anisotropy (see \sref{SS-inv}).
}
\label{fig2}
\end{figure*}

The labels a, b, c, d, e, and f in \fref{fig1} indicate the slit
position of the six profiles shown in \fref{fig2}, each profile plotted
using the same color as the corresponding label. The profiles at locations 
b, c, and d are also shown in \cite{Ishikawa2021}. 
The profiles in panels (A) and (B) in \fref{fig2} (a,b, and c) are located in
the plage region, while those in (C) and (D) are located in the enhanced
network (d and e) and quiet Sun (f) regions.

The intensity profiles corresponding to the plage (panel A of
\fref{fig2}) are significantly brighter and show a relatively
shallow central trough compared to the profiles outside the plage
(panel C). Regarding the circular polarization profiles, the
ones in the plage (panel B of \fref{fig2}) show the typical 
spectral shape with four lobes, while those outside the 
plage show significantly depressed inner lobes which are almost absent.

The inverted model atmospheres in the plage region are similar to those inferred by 
\cite{delaCruz2016ApJ}, showing higher temperature
(panel E of \fref{fig1} and \fref{fig2})
and gas pressure (panel H of \fref{fig1}) compared to the region
outside 
the plage. Moreover, the temperature in the lower
chromosphere (from \TAUA{-4} to $\approx-5.3$) increases to more than
$6000$~K, a temperature enhancement when compared to the models
from outside the plage. A similar result was found by
\cite{Carlsson2015ApJ}, who introduced a chromospheric plateau of about $6500$~K
in model P of \cite{Fontenla1993} to be able to
fit some plage observations of the \ion{Mg}{2} h and k lines from the
\emph{Interface Region Imaging Spectrograph}
\citep[IRIS;][]{DePontieu2014}.

The three selected plage profiles have k$_{2v} >$ k$_{2r}$ and
h$_{2v} >$ h$_{2r}$, which is associated with negative velocity
gradients in the middle and upper chromosphere
\citep[i.e., macroscopic velocity decreasing with height; see also][]
{Leenaarts2013ApJ,Afonso2023ApJ}. The inverted {\vlos} (panel
G of \fref{fig2}) confirms this
expected behavior. On the contrary, the profile at location e
(see the gray curves in panel C of \fref{fig2})
has k$_{2r} >$ k$_{2v}$ and
h$_{2r} >$ h$_{2v}$, indicative of positive velocity gradients,
as recovered by the inversion (gray curve in panel H of
\fref{fig2}).

A larger {\vmi} in the lower chromosphere is necessary to fit the
relatively broad wings close to the k$_3$ and h$_3$ 
(compared to the rest of the selected
profiles) of the plage profiles at locations a and c (panel I
of \fref{fig2}).
The narrower peak separations in these two profiles (a and c) 
result, in turn, in smaller {\vmi} in the middle chromosphere
with respect to the other four selected profiles.

Finally, we show the inverted {\blos} in panels K and L of \fref{fig2},
which correspond to slit positions inside and outside the bright 
region of the observed plage, respectively.
The solid curves show the result of the inversion taking into account the impact
of radiation anisotropy (fifth inversion cycle), while
the dashed curves show the result of the inversion 
by neglecting it (fourth inversion cycle). Neglecting the
impact of radiation anisotropy results in the retrieval of
larger magnetic fields in the plage region, as explained in
\cite{Li2022}. Although the anisotropy only impacts the outer lobes, 
and thus we should expect this difference to appear just in the lower
chromosphere, we see an effect on the inverted {\blos} in the upper
chromosphere due to the coupling induced by the spectral \gls*{psf}
\citep{Centeno2022ApJ}.

In the three plage locations (a, b, and c) we get {\blos} of up to $300$~G
in the middle chromosphere, decreasing with height. 
This magnetic field strength is of similar order to those
reported by \citet{Morosin2020A&A} and \citet{Pietrow2020A&A}
in plage regions, obtained from the inversion of spectropolarimetric
data in the \ion{Ca}{2} line at $854.2$~nm.
In the enhanced network position (d) {\blos} is about $88$~G at \TAUE{-4.38}, around
$10$\% larger than what was estimated by \cite{Ishikawa2021} by
applying the \gls*{wfa} to the outer lobes of the h line circular
polarization profile. Note that the application of the \gls*{wfa} to
the outer lobes of the \ion{Mg}{2} h-k doublet circular polarization
profiles generally underestimates {\blos} and that this error is
of the order of $10$\% for the h line \citep{Ishikawa2021}.
For this enhanced
network location the magnetic field in the upper chromosphere is close
to zero, as expected from the very small inner lobes in the circular
polarization profile. The fit to these profiles (locations f
and e) are worse. In these quieter regions of the Sun we expect magnetic
concentration of smaller scale \citep{Stenflo1989A&ARv} and, thus, not only
the magnetic field is probably not filling the full pixel of the
observation, but several different longitudinal magnetic fields can
coexist in each pixel leading to significant signal cancellation.

%###############################################################################

\subsection{Comparison with previous results}\label{SS-comp}

\cite{Ishikawa2021} first analyzed the data used in this work by applying the
\gls*{wfa} to infer {\blos} at different heights in the solar
atmosphere. The discrete heights of the {\blos} values inferred via the \gls*{wfa}
are only approximate, with the locations along the LOS based on
theoretical investigations \citep{Afonso2023ApJ,Tanausu2022ApJ}.
In this work we have instead applied an inversion code and we are thus able to
infer the whole stratification of those regions contributing to the formation
of the emergent Stokes parameters. 
Despite these differences, it is valuable
to compare the two results and to determine the degree of agreement between them.

\begin{figure*}[htp]
\center
\includegraphics[width=0.70\textwidth]{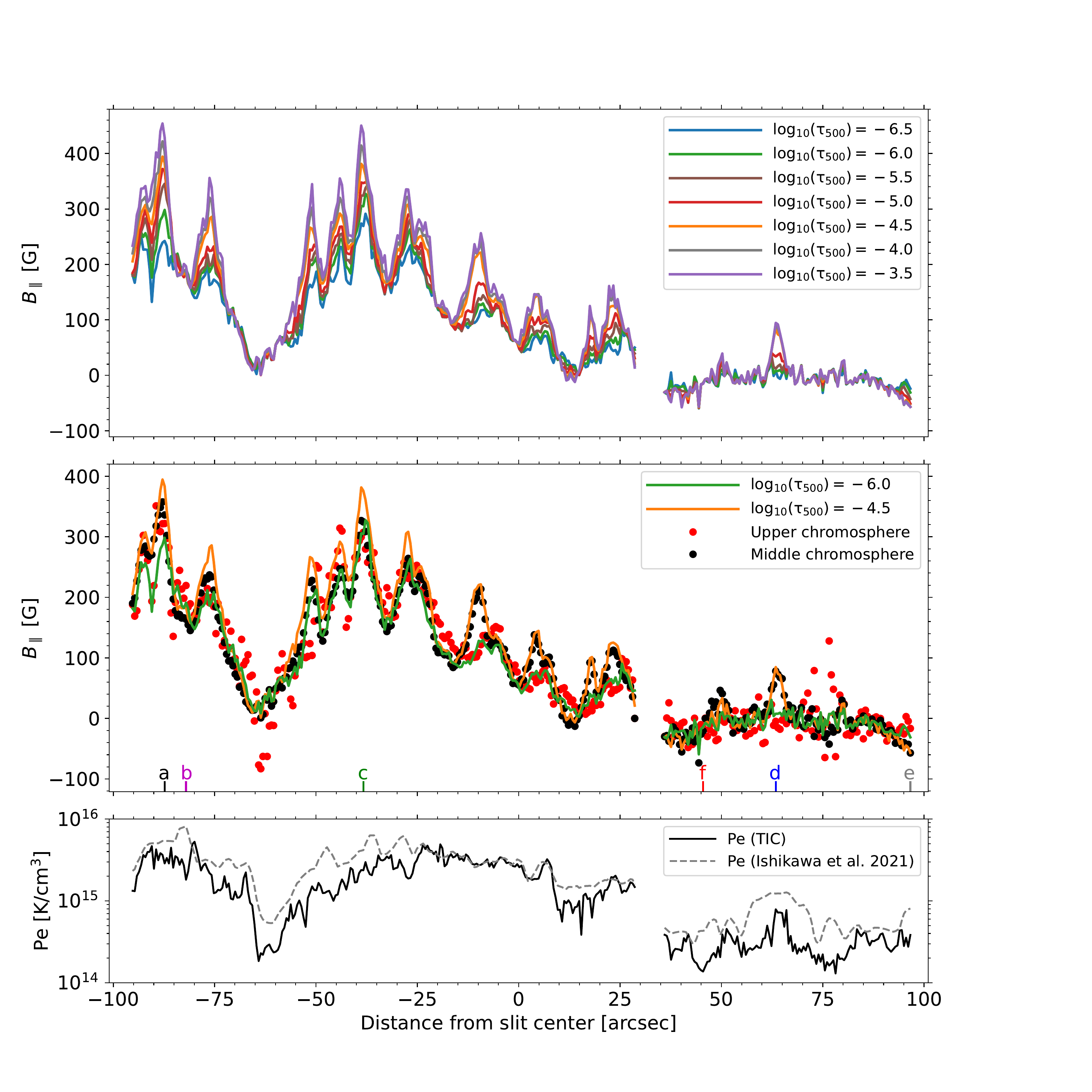} \\
\caption{Upper panel: inverted {\blos} at \TAUE{-6.5, -6.0, -5.5, -5.0, -4.5, -4.0}, and $-3.5$ 
(solid curves, see legend) for each slit position. 
Middle panel: inverted {\blos} at \TAUE{-6.0} and $-4.5$.
The filled circles show the {\blos} in Fig.~3 of \cite{Ishikawa2021}, with the same
color legend, namely, red for upper chromosphere and black for middle chromosphere. 
Bottom panel:  the black solid curve shows the inverted electron pressure at \TAUE{-5.4}
(see right axis) for each slit position. 
The gray dashed curve shows the electron pressure at \TAUE{-5.4} from \cite{Ishikawa2021}.
}
\label{fig3}
\end{figure*}

In the upper panel of \fref{fig3},  we show the inferred {\blos} at every position of the slit at 
various heights. The {\blos} values in the lower chromosphere ($\sim$\TAUE{-4.0}) and 
at the temperature
minimum ($\sim$\TAUE{-3.5}) are somewhat smaller than those inferred from the \ion{Mn}{1} lines
in \cite{Ishikawa2021}. This is due to the lack of sensitivity of the \ion{Mg}{2} lines
to these layers of the model atmosphere. 
In order to get a more reliable {\blos} estimation in the region just above the 
temperature minimum using the CLASP2 data, we should include the inversion of the Mn I lines. 
This, however, is currently beyond the capabilities of TIC, as it requires to take into account the 
hyperfine structure of Mn I  \citep{Tanausu2022ApJ}.
The inverted {\blos} are similar to the results
of \cite{Ishikawa2021} for the middle and upper chromosphere. 
Because the particular height that should be assigned to the {\blos} retrieved by applying the WFA depends on the particular
stratification of the atmosphere and the source function for the circular polarization,
we cannot expect to find a particular optical depth that perfectly matches
the \gls*{wfa} inference everywhere. Nevertheless, we can see that the {\blos} values inferred
for the middle chromosphere in are mostly between our \TAUE{-4.5} and
$-5.0$ values, and that those for the upper chromosphere are mostly between our
\TAUE{-5.5} and $-6.0$ values. We note that the two inferences agree within 
the error bars shown in  \citet{Ishikawa2021} and in our Fig 2.

The black solid curve in the bottom panel of \fref{fig3} shows the TIC-inferred electron pressure, 
i.e., the product of the temperature and the electron density, at \TAUE{-5.4}. 
The gray dashed curve shows the electron pressure obtained by
\cite{Ishikawa2021} at the same optical depth inverted with IRIS$^2$ \citep{Sainz2019}. 
While the overall trend is similar, the two determinations of electron pressure show clear
differences. This is to be expected, since both 
codes only invert for the electron density at the top boundary, while the stratification of this quantity 
relies on the assumption of hydrostatic equilibrium, and the treatment of the equation of 
state by the two codes is different. However, both inversions demonstrate that the electron pressure 
correlates with the inverted {\blosp}, supporting the magnetic origin of the 
upper-chromosphere heating above active region plages. 
In particular, the TIC inversion gives Pearson correlation coefficients
of 0.92, 0.68, and 0.74 between {\blos} and $P_{\rm e}$ at \TAUE{-6.0}, $-5.4$, and
$-5.0$, respectively.

In conclusion, our results on the stratification of {\blosp} obtained by applying 
TIC to the CLASP2 plage data agree with the results of \cite{Ishikawa2021}, 
obtained by applying the \gls*{wfa} to the same data, and assigning the inferred value of {\blosp} 
to regions in the solar atmosphere based on
theoretical studies of the formation of the \ion{Mg}{2} h-k doublet.

%###############################################################################

\subsection{Magnetic energy in the plage chromosphere}\label{SS-ene}

\begin{figure*}[htp]
\center
\includegraphics[width=1.00\textwidth]{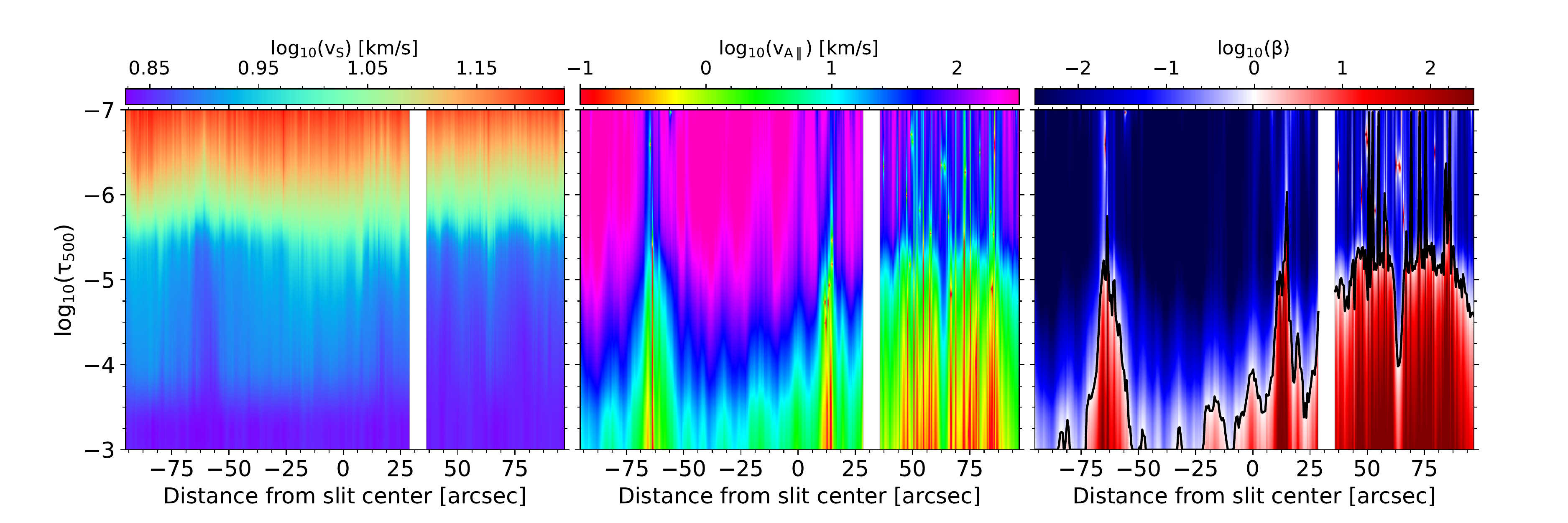}
\caption{Speed of sound (left panel), Alfv\'en velocity (middle panel), and ratio 
between gas pressure and the magnetic
pressure (right panel), with the black
solid curve indicating where this ratio is unity. The two middle and right panels 
assume that the field is fully longitudinal, i.e., $B=${\blos}}
\label{fig4}
\end{figure*}

\begin{figure*}[htp]
\center
\includegraphics[width=0.90\textwidth]{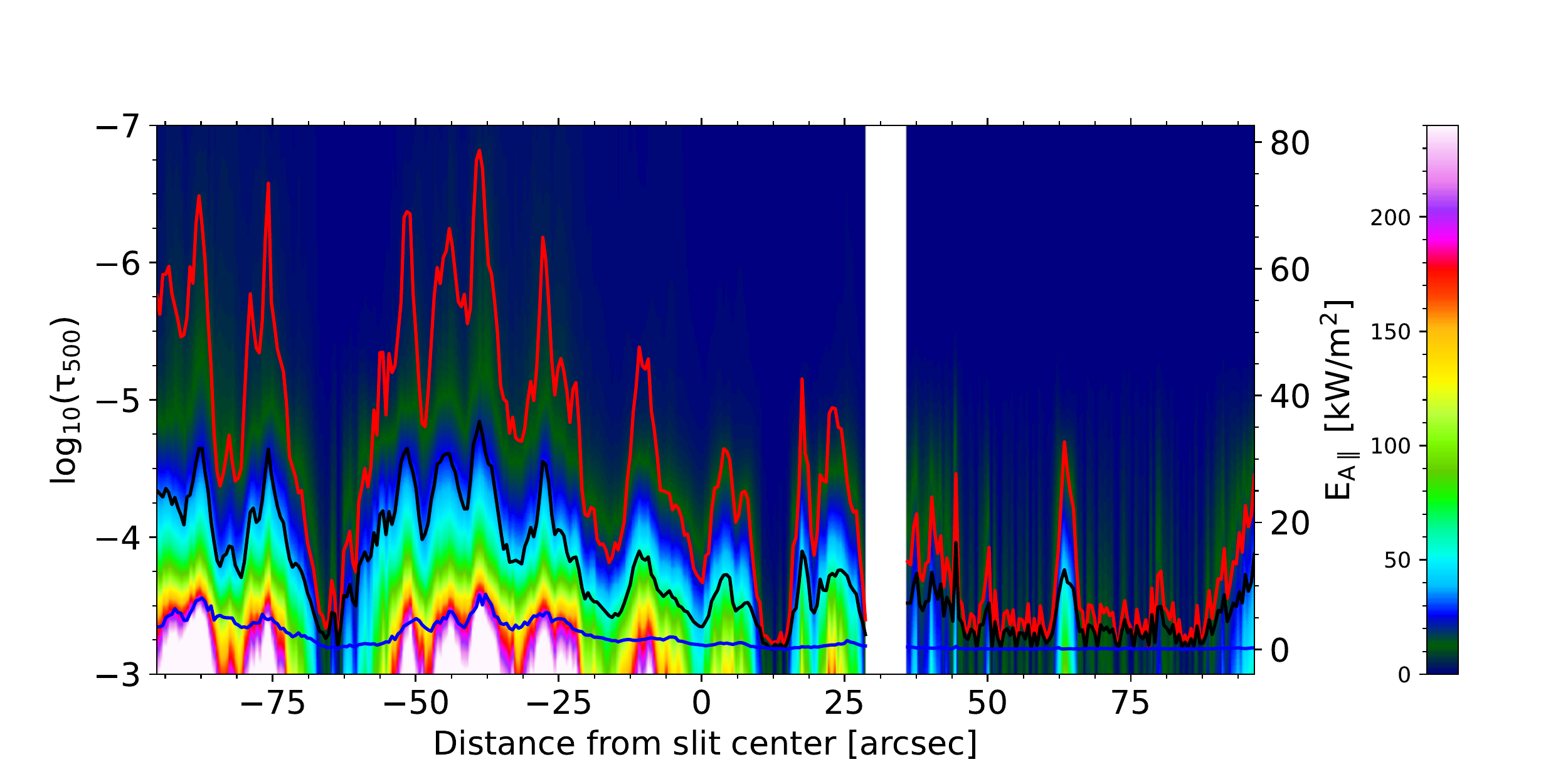}
\caption{Energy flux of Alfv{\'e}n waves by assuming that the magnetic field is
fully given by \blos and that the velocity amplitude is $v=2$~km/s. 
The red, black, and blue curves denote the energy flux at \TAUE{-4.0, -4.5}, and $-5.8$, respectively.}
\label{fig5}
\end{figure*}

We used the inferred model atmosphere described in the previous sections to estimate
the speed of sound ($v_s=\sqrt{\gamma P_{\rm g}/\rho}$, where $\gamma$ is the
adiabatic index that we take as $\gamma=5/3$ and $\rho$ is the density) and
the Alfv{\'e}n velocity (under the assumption that the magnetic field is
fully directed along the \gls*{los};
$v_{A}=\sqrt{B^2_{\parallel}/(\mu_0 \rho)}$, in SI units, \citealt{Priest2014},
where $\mu_0$ is the vacuum permeability) in the chromosphere of the observed plage region. 
We show these two quantities in the left
and middle panels of \fref{fig4}, respectively. 
The right panel shows the plasma $\beta=2\mu_0 P_g/B^2$\textemdash i.e., 
the ratio between the gas and magnetic pressures\textemdash  assuming 
that the magnetic field is fully along the LOS 
(i.e., $B=B_\parallel$). The $\beta=1$ surface is indicated
by the black curve in the corresponding panel. Around the slit positions at $-90$, $-75$, 
$-40$, and $-25$ ~arcsec the magnetic pressure has 
overcome the gas pressure ($\beta < 1$) already at \TAUE{-3}. In the rest of the plage, we find $\beta=1$
in the lower chromosphere, except around the slit position around $-60$~arcsec, where
the magnetic field drops to very small values, and close to the border of the plage
region around the slit position at $10$~arcsec. Outside of the plage, the $\beta=1$
surface is in the middle chromosphere instead, except in the enhanced network
region where the $\beta=1$ surface is lower than in the surrounding quiet regions.
The shape of the $\beta=1$
surface inside of the plage resembles that of upside-down ``cones'', what may be
indicative of magnetic flux concentrations expanding with ``height''.\footnote{Note that
we sample the atmospheric models in optical depth and therefore two positions with
the same optical depth in two different columns are not necessarily one next to
the other in the geometrical scale. However, if the thermodynamic quantities
are similar between neighbor columns we can expect the corresponding geometrical
heights to also be relatively similar.}

Several different mechanisms have been proposed for the heating of the chromosphere above 
plages, e.g., magneto-acoustic shocks, Alfv\'en waves, magnetic reconnection, and 
Ohmic dissipation \citep[e.g.,][]{Carlsson2019ARA&A,Anan2021ApJ}.
Certainly, the strong correlation between the inferred {\blos} and the electron pressure 
discussed in the previous section supports the hypothesis that chromospheric heating 
above plages is of magnetic origin.
In particular, Alfv\'en waves have been
considered to play an important role in chromospheric heating and they have been
observed in chromospheric structures
(\citealt{DePontieu2007Sci,Jess2009Sci,Jess2015SSRv,Grant2018NatPh}, and
\citealt{VanDoorsselaere2020SSRv}). From \gls*{mhd} simulations,
\citet{vanBallegooijen2011ApJ} pose that Alfv\'en waves produced by the motion
of the footpoint of a flux tube can be reflected due to the steep increase of
the Alfv\'en speed with height. The interaction between the counterpropagating
waves lead to the Alfv\'en wave turbulence and dissipation in the chromosphere.
\cite{Soler2017ApJ} studied the propagation of torsional Alfv\'en waves in expanding
flux tubes, finding that only the lower frequency waves are reflected, while
the high frequency ones are damped in the chromosphere by ion-neutral collisions,
with only the intermediate frequency waves being transmitted into the corona. 
Our inversion results show significant {\vmi} in the lower chromosphere of the
plage region (see panel G of \fref{fig1} and curves a and c in panel G of
\fref{fig2}), in agreement with the inversions of \cite{delaCruz2016ApJ} in
IRIS observations. {\vmi} accounts for many broadening contributions that cannot
be accounted for in 1D radiative transfer modeling.
Therefore, it is not possible to establish a direct
relation between the increased {\vmi} in the plage region inversion with the presence
of this Alfv\'en wave turbulence. Nevertheless, it is worth noting that it is a
plausible explanation of the results.

The total duration of the CLASP2 observation is too short for investigating whether or not 
the observed Stokes profiles show any hint of Alfv\'en waves.
However, torsional Alfv\'en waves have been observed in the
network with velocity amplitudes of $2.6$~km/s \citep[e.g.,][]{Jess2009Sci}. By assuming 
that Alfv\'en waves do exist in the observed plage region and that their
amplitude is about $v=2$~km/s we can have a rough estimation of the energy
flux carried by Alfv\'en waves ($E_A$), also assuming that the
magnetic field is fully given by \blos ($E_{A}=\rho v^2 v_{A}$).

In \fref{fig5} we show this rough estimate of the energy flux that could be
carried by Alfv\'en waves, with the red, black, and blue curves showing
the energy flux (right axis of the figure) corresponding to \TAUE{-4.0},
$-4.5$, and $-5.8$, respectively. At \TAUE{-4.0} the energy flux can
reach 60~kW/m$^2$ inside the plage, while it is around 30~kW/m$^2$ in
the network. This value is of similar order than the energy loss 
estimated by \citeauthor{Withbroe1977ARA&A} (\citeyear{Withbroe1977ARA&A};
$2\cdot10^7$~erg/cm$^2$/s, i.e. $20$~kW/m$^2$) 
and by \citeauthor{Morosin2022A&A} (\citeyear{Morosin2022A&A}; $26.1$~kW/m$^2$)
in the chromosphere of active regions.
At \TAUE{-4.5} the energy flux would be around $30$~kW/m$^2$ in the 
plage and around $10$~kW/m$^2$ in the network.
Therefore, Alfv\'en waves with a velocity amplitude of $2$~km/s could
carry enough energy to balance out the energy losses in the chromosphere
in the plage region. It is important to take into account that we are
assuming that the magnetic field strength is given by {\blosp}, 
and this is a lower limit of the actual field strength. Therefore, Alfv\'en amplitudes 
even smaller than $2$~km/s may already be sufficient to reach this energy flux.

%###############################################################################
%###############################################################################
%###############################################################################

\section{Summary and discussion}\label{S-conclusions}

We have applied the TIC to the intensity and circular polarization profiles observed 
by the CLASP2 in an active region plage and enhanced network regions, obtaining the stratification
of the temperature (T), line of sight velocity ({\vlosp}), microturbulent
velocity ({\vmip}), gas pressure (\Pgp), and the longitudinal component of the magnetic field
 ({\blosp}) at each position along the spectrograph slit. In the
inversion we have fit the intensity and circular polarization profiles of the
\ion{Mg}{2} h-k doublet and subordinate triplet at 279.88~nm.
The general properties of the atmospheric models obtained are in agreement 
with the results presented in previous works. The
inverted plage models show larger temperature and gas pressure,
similar to those obtained by \cite{delaCruz2016ApJ}
for a plage observation from IRIS. Likewise, the plage models obtained show a 
plateau-like region in the temperature stratification with values around 6500 K, 
which is the same temperature of the plateau that \cite{Carlsson2015ApJ} added to the
P model of \cite{Fontenla1993} in order to fit some plage observations from
IRIS.

The longitudinal magnetic field that results from the inversion is in agreement with the \gls*{wfa}
inference by \cite{Ishikawa2021} for the middle and upper chromosphere,
especially when taking into account that the application of the \gls*{wfa} 
to the wings of the \ion{Mg}{2} h line
systematically underestimates the magnetic field by about 10\%. Our inversions
retrieve smaller \blos in the lower chromosphere when compared with the results of 
\cite{Ishikawa2021}. However, this may be due
to a lack of sensitivity to those layers of the atmosphere (see the circular
polarization response function in Fig.~3 of \citealt{Tanausu2022ApJ}). The
inversion at these atmospheric layers could be significantly improved by
including the \ion{Mn}{1} lines, however, being \ion{Mn}{1} a minority species
(which usually entails relatively large atomic models to correctly model their
spectral lines) and the need to account for hyperfine structure (especially
to correctly model their circular polarization profiles) would entail
a prohibitive increase of the required computing time. 

In the upper chromosphere, the electron pressure derived 
with TIC is not identical to that shown in \cite{Ishikawa2021} derived
with IRIS$^2$. However, this is expected considering the 
different treatment of the equation of state by the two codes. Nevertheless, what is really
important is that there is agreement in the correlation between the electron
pressure and the {\blosp}, which strongly supports the magnetic origin of the
heating of the upper chromosphere.

From the inverted model atmospheres we have estimated the Alfv\'en velocity
and the plasma $\beta$ under the assumption that the magnetic field is
fully given by {\blosp}. The shape of the $\beta=1$ surface in the plage, namely, 
that of inverted ``cones'', may be indicative of magnetic flux concentrations expanding
with height. We need to take into consideration, however, that our inversions
are in optical depth scale, and two positions at the same optical depth in
two different columns are not necessarily one next to the other in geometrical
scale. Nevertheless, if the thermodynamic quantities are similar between
neighbor columns, we can expect adjacent points in optical depth to be
relatively close in geometrical height as well.

Finally, by assuming a velocity amplitude for Alfv\'en waves of $2$~km/s, based
on the previous results in the literature \citep{Jess2009Sci}, we have estimated
the energy flux carried by Alfv\'en waves in the inferred model atmosphere. The energy
flux inside the plage at the middle chromosphere (\TAUE{-4.5}) is about
30~kW/m$^2$, of similar order as the energy loss of $20$~kW/m$^2$ in the chromosphere of
active regions estimated by \cite{Withbroe1977ARA&A}. 
At \TAUE{-4.0}, this energy flux reaches 60~kW/m$^2$. 
Consequently, Alfv\'en waves with a velocity amplitude of $2$~km/s
could carry enough energy to balance out the energy losses in the chromosphere
of the plage region. Moreover, our assumption of the magnetic field strength
being given uniquely by {\blos} can only be a lower limit of the actual field
strength and, therefore, smaller velocity amplitudes 
may be sufficient in order to reach these energy fluxes.

As shown in this paper, the TIC is a very useful plasma diagnostic tool for inferring the magnetic, 
thermodynamic and dynamic properties of the solar chromosphere from spectropolarimetric 
observations of the  
resonance lines, such as the \ion{Mg}{2} h and k doublet, that the CLASP2 suborbital
space experiment has made possible. 
We therefore believe that the diagnostic capabilities of the scientific community 
are ready to be applied to a mission dedicated to these observables, and that 
the development of a space telescope equipped with instruments like CLASP2 would lead to a 
revolution in our empirical understanding of the magnetic field in the solar upper atmosphere.

%###############################################################################
%###############################################################################
%###############################################################################

\acknowledgements
We thank the referee for a careful reading of the manuscript and the helpful suggestions and comments.
We acknowledge the funding received from the European Research Council (ERC)
under the European Union's Horizon 2020 research and innovation programme
(ERC Advanced Grant agreement No 742265). 
CLASP2 is an international partnership between NASA/MSFC, NAOJ, JAXA, IAC, and IAS; 
additional partners include ASCR, IRSOL, LMSAL, and the University of Oslo.
The Japanese participation was funded by JAXA as a Small Mission-of-Opportunity Program, 
JSPS KAKENHI Grant numbers JP25220703 and JP16H03963, 2015 ISAS Grant for Promoting 
International Mission Collaboration, and by 2016 NAOJ Grant for Development Collaboration. 
The USA participation was funded by NASA Award 16-HTIDS16\_2-0027. 
The Spanish participation was funded by the European Research Council (ERC) under the 
European Union's Horizon 2020 research and innovation programme (Advanced Grant agreement No. 742265). 
The French hardware participation was funded by CNES funds CLASP2-13616A and 13617A.

%###############################################################################
%###############################################################################
%###############################################################################

\appendix
\counterwithin{figure}{section}

\section{fit profiles for the whole spectral range}

\begin{figure*}[htp]
\center
\includegraphics[width=1.0\textwidth]{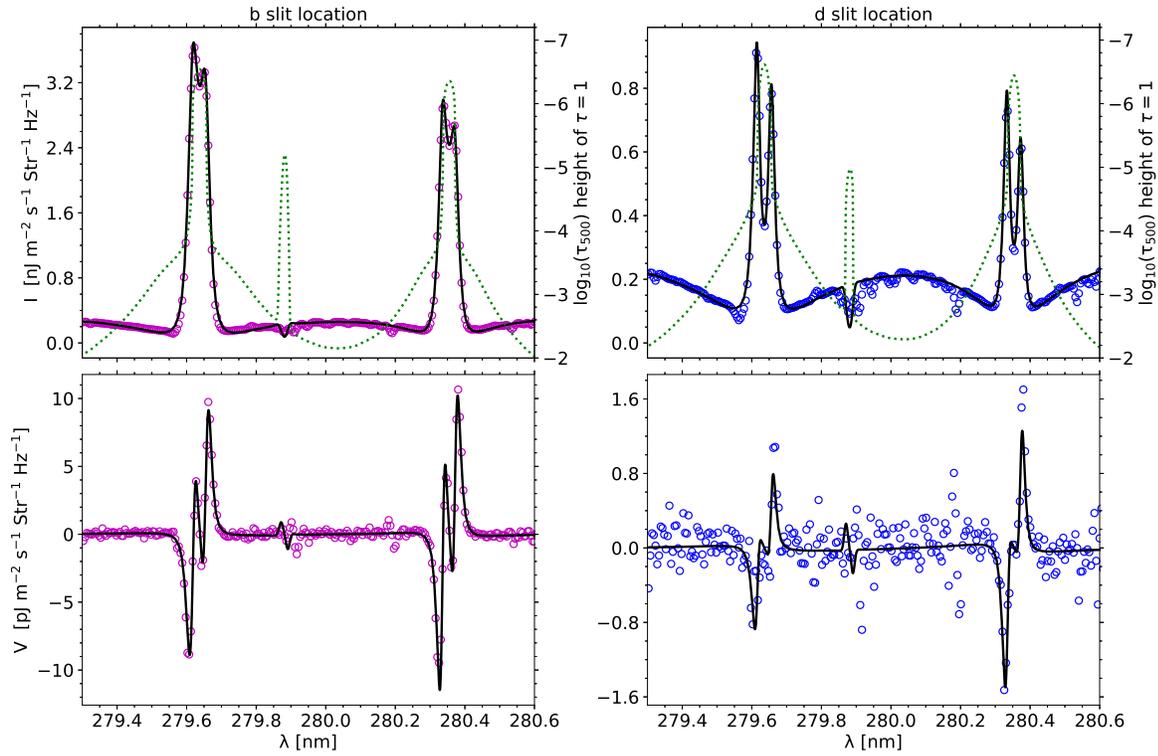}
\caption{Observed (open circles) and fit (solid curves) intensity (top panels)
and circular polarization (bottom panels) profiles for a position in
the plage (b; left panels) and for a position in the enhanced network (d; right panels).
For more information see Fig. 2. The green dotted curves show, for each wavelength,
the optical depth (in ${\rm log_{10}}(\tau_{500})$) where $\tau=1$.
}
\label{figa1}
\end{figure*}

In \fref{figa1} we show the fit to the intensity and circular polarization profiles
corresponding to the b and d locations in the slit (purple and blue curves in \fref{fig2},
respectively) for the whole CLASP2 spectral range, including the subordinate
\ion{Mg}{2} lines between the h and k lines. In the inversion we exclude the wavelengths
where there are spectral lines other than the \ion{Mg}{2} k-h doublet and its subordinate
triplet.

\bibliography{plage}{}

\bibliographystyle{aasjournal}

\end{document}